\documentclass[ams]{article}

\def\alf{\alpha}     
\def\veps{\varepsilon}

\def\la{\langle}   \def\ra{\rangle}
   \def\dg{\dagger}

\def\noi{\noindent}   
\def\dsty{\displaystyle}
\def\uln{\underline}

\def\beq{\begin{equation}}
\def\eeq{\end{equation}}
\def\bea{\begin{eqnarray}}
\def\eea{\end{eqnarray}}

\def\bt{\begin{tabular}}
\def\et{\end{tabular}}

\def\lb{\label}
\begin{document}

\title{\bf Nonlinear Fermions and Coherent States}
\author{D.A. Trifonov\\
Institute for Nuclear Research and Nuclear Energetics,\\
 72 Tzarigradsko chaussee, Sofia, Bulgaria}
\maketitle

\begin{abstract}
Nonlinear fermions of degree $n$ ($n$-fermions) are introduced as particles with  creation and annihilation operators obeying the simple nonlinear anticommutation relation $AA^\dagger + {A^\dagger}^n A^n =1$. The ($n+1$)-order nilpotency of these operators follows from the existence of unique $A$-vacuum. Supposing  appropreate ($n+1$)-order  nilpotent  para-Grassmann variables and integration rules the sets of $n$-fermion number states, 'right' and 'left' ladder operator coherent states   (CS) and displacement-operator-like CS   are constructed.    The $(n+1)\times(n+1)$ matrix realization of the related para-Grassmann algebra is provided.   General $(n+1)$-order nilpotent ladder operators of finite dimensional systems are expressed as polynomials in terms of $n$-fermion operators. Overcomplete sets of  (normalized)  'right' and 'left' eigenstates of such general ladder operators  are constructed and their properties briefly discussed.  

\end{abstract}

\section{Introduction}

In the last decades or so a considerable attention is paid in literature to the problems of extension and adoption of the celebrated coherent state (CS) method \cite{Glauber, Klauder, Ali} for description of quantum systems with finite dimensional Hilbert state space - fermionic \cite{Abe, Cahill, Grigorescu}, parafermionic ($k$-fermionic) \cite{Daoud'02, Baz'02, Cabra'06, Baz'10} and parabosonic \cite{Chakrabarti}, Hermitian and pseudo-Hermitian \cite{Cherbal'07, Cherbal'10, Fasihi'10, N-bashi'11, Maleki'11}, systems with discrete finite coordinate spectrums  \cite{Miranowicz, Vourdas}. Finite-dimensional quantum mechanics proved useful in many areas, such as quantum computing, quantum optics, signal analysis etc. (see e.g. \cite{Vourdas, Jafarov} and references therein). 

Parafermions and parabosons were introduced by Green \cite{Green, Messiah} in order to study particle statistics of a more general type than the common Fermi-Dirac and Bose-Einstein statistics. The annihilation and creation operators of these (hypothetical) particles obey certain trilinear commutation relations  for parafermions and commutation-anticommutation relations for parabosons. 
The maximal number of parafermions in a given state is finite, denoted usually as $p$,  and called the order of (para)statistics. These ladder operators are nilpotent of order $p+1$.  It was later shown \cite{Ryan} that $n$ pairs of Green parafermions generate Lie algebra of the orthogonal group $SO(2n+1)$.  Palev \cite{Palev} introduced creation and annihilation operators, $n$ pairs of which generate the algebra of the unimodular group $SL(n+1)$, the corresponding order of statistics  $p$ (called $A_n$-statistics) being also finite. The dimension of $n$-mode Palev parafermion Fock space is $(p+n)!/p!n!$, so the ladder operators are nilpotent of order $p+1$. CS  of Klauder-Perelomov type for these parafermions are constructed by Daoud \cite{Daoud'06}. For Green parabosons the ladder operator CS (overcomplete in a certain subspace)  are constructed for the  two mode case in \cite{Chakrabarti}. 

Finite dimensional Fock spaces can be constructed by means of ladder operators that obey the $q$-deformed boson commutation relations \cite{Biedenharn} when $q$ is $k$-root of unity, the corresponding parafermions being called $k$-fermions \cite{Daoud'02}. Ladder operator CS for Hermitian $k$-fermions are considered in \cite{Daoud'02, Cabra'06, Baz'10, N-bashi'11}, for pseudo-Hermitian - in \cite{Fasihi'10,  Maleki'11}.       These CS are constructed as non-normalized  'left' eigenstates of the corresponding ladder operators, but when normalized they cease being eigenstates of that operators due to noncommutation between paragrassmanian eigenvalues and the normalization factors. Another unsatisfactory feature of such ladder operator parafermion CS is that the eigenvalue of the squared ladder operator $b^2$ is not equal to the square of the eigenvalue $\theta$ of $b$, which is due to the noncommutation between $\theta$ and  $b$.  The  displacement-operator CS are also lacking  . 
  
In the present paper we introduce new kind of parafermions,  based on simple non-linear anticommutation relations (suggested  by Chaichian and Demichev \cite{Demichev} in the context of polynomial relations for the generators of the su$(2)$ Lie algebra) and construct the related Fock states, CS   in the form  normalized 'left' and 'right' ladder operator eigenstates,  and displacement-operator-like CS  .    We call these parafermions $n$-linear fermions (shortly $n$-fermions), where the positive integer $n$ is the degree of nonlinearity of  anticommutation relations. A remarkable feature of $n$-fermions is that the order of their statistics equals the degree of nonlinearity $n$.      An advantage of the 'right' CS is that they are free from the above noted unsatisfactory features of 'left' parafermion CS.     The related generalized Grassmann variables   and their integration rules that ensure the overcompleteness of $n$-fermion CS, appear to be direct and most simple extension of the standard fermion variables and integration rules.    They  ensure the resolution of the identity in terms of  $n$-fermion ladder operator CS with no additional weight functions, unlike the case of previous parafermion CS \cite{Daoud'02, Cabra'06, Baz'10, Fasihi'10, N-bashi'11}, where the introduction of weight functions is always needed. We had to introduce weight functions in the case of the displacement-operator-like $n$-fermion CS only. The  algebra of our ($n+1$)-order nilpotent para-Grassmann variables admits simple $(n+1)\times(n+1)$ matrix realization  .  

The organization of the paper is as follows. In the second section $n$-fermion algebra and Fock states are considered. Supposing the existence of $n$-fermion vacuum the ($n+1$)-order nilpotency of creation and annihilation operators is derived and $n+1$ excited number states are constructed. 
In the third section normalized 'right' eigenstates of the annihilation operator are built up and their overcompleteness is established, using appropriately defined anticommutation relations between 
($n+1$)-order nilpotent eigenvalues $\zeta$, $\zeta^*$ and $n$-fermion ladder operators, and new integration rules for the  variables $\zeta$, $\zeta^*$. 
The structure of the integration rules is different for even and odd $n$, being most simple for odd $n$ (even dimension of the Fock space). In the fourth section we consider general form of ladder operators in finite dimensional Hilbert space (finite level systems) and construct the related normalized 'right' CS. The form of the weight function that ensure the identity resolution of the general ladder operator CS with respect to the same integration rules, as introduced for the $n$-fermion case, is explicitly determined. The general ladder operators for finite level systems are expressed as polynomials in terms of the $n$-fermion creation and annihilation operators. In the last section the nonnormalized 'left' general ladder operator eigenstates     and displacement-operator-like CS      are constructed, and their overcompleteness established. Some objectionable properties of  'left' CS are pointed out      and the corresponding advantages of the 'right'  CS are noted  .

\section{Nonlinear fermion algebra and Fock states}

Consider the  operators $A(n)$ and $A^\dg(n)$ which satisfy the nonlinear anticommutation relation of the form\footnote{Such  relation  has been suggested (and realized for $n=2,3$) in ref.  \cite{Demichev} in the context of polynomial relations for the generators of the su$(2)$ Lie algebra.} 
\begin{equation}\label{{A,A^dg}}
A(n)A^\dg(n) + {A^\dg}^n(n)A^{n}(n) = 1,
\end{equation} 
$n$ being a positive integer.
At $n=1$ the standard fermionic relations $aa^\dg + a^\dg a = 1$ are recovered, i.e.  $A(1)= a$. 
Therefore the nonlinear relation (\ref{{A,A^dg}}) could be called {\it nonlinear fermion anticommutation relation}.  
In analogy to the case of standard fermions we are aimed to interpret operators  $A(n)$ and $A^\dg(n)$ as particle  annihilation and creation operators, and $n$ as the highest number of particles in one quantum state.   In this aim we have to construct Fock orthonormal states  and a relevant number operator. 
The related particles could be called  {\it nonlinear fermions} with  degree of nonlinearity $n$ and order of statistics\cite{Green, Messiah} $n$ (order of nilpotency $n+1$), shortly $n$-{\it linear fermions}, or simply {\it $n$-fermions}. 
In this terminology the standard fermions are "$1$-fermions", or {\it linear fermions}.  

We suppose that there exists a vacuum state $|0\ra$ that is annihilated by $A(n)$, 
\begin{equation}\label{A|0>}
A(n)|0\ra = 0.  
\end{equation}
From the uniqueness of the vacuum one can derive that  $A(n)$ is nilpotent of order $n$, namely $A^{n+1}(n)=0$. In this aim we first consider $A(n){A^\dg}^{n+1}(n)$. Applying twice the anticommutation (\ref{{A,A^dg}}) to $A(n){A^\dg}^{n+1}(n)$ we have 
\beq\lb{*0}
\bt{l}
$A(n){A^\dg}^{n+1}(n) = {A^\dg}^n(n) - {A^\dg}^n(n) A^n(n) {A^\dg}^n(n)$ \\ $= {A^\dg}^n(n) - {A^\dg}^n(n)  A^{n-1}(n) {A^\dg}^{n-1}(n)  + {A^\dg}^n(n) A^{n-1}(n) {A^\dg}^n(n) A(n) A^{n-1}(n){A^\dg}^{n-1}(n).$
\et \eeq
The last two terms in (\ref{*0}) contain factor $A^{n-1}(n) {A^\dg}^{n-1}(n)$ to the right, which by repeated applications of (\ref{{A,A^dg}}) can be transformed to  the form 
\beq\lb{*1}
A^{n-1}(n) {A^\dg}^{n-1}(n) = 1 \, + \,{\rm  terms\,\, with}\,\, A^{i'}(n), \,i'\geq n-1, \,\,{\rm to \,\,the\,\, right}.
\eeq
Taking this into account we see that all terms in r.h.s. of (\ref{*0}) contain factor $A(n)$ to their right, which means that the operator $A(n){A^\dg}^{n+1}(n)$ annihilates $|0\ra$, and thus $A(n)$ annihilates the state ${A^\dg}^{n+1}(n)|0\ra$.  In view of the uniqueness of $|0\ra$ we have to put ${A^\dg}^{n+1}(n)|0\ra = c_1|0\ra$. Multiplying the latter equation from the left by $\la 0|$ we get $c_1=0$. This proves that  ${A^\dg}^{n+1}(n)$ annihilates $|0\ra$,
\beq\lb{*2}
{A^\dg}^{n+1}(n)|0\ra  = 0.
\eeq
Next we have to consider the states $|k\ra$,
\begin{equation}\label{|k>}
|k\ra = {A^\dg}^k(n)|0\ra.  
\end{equation}
In view of (\ref{*2}) there are $n+1$ nonvanishing states $|k\ra$, $k=0,1, \ldots, n$. These $|k\ra$ can be proved to be orthonormalized,
\begin{equation}\label{<i|k>}
\la i | k \ra =  \la 0|A^i(n) {A^\dg}^k(n)|0\ra =   \delta_{ik}.  
\end{equation}
To   verify (\ref{<i|k>}) we consider the products $A^i(n){A^\dg}^k(n)$. Applying repeatedly (\ref{{A,A^dg}}) we find, similarly to the case (\ref{*1}),    
\beq\lb{*3}
\bt{l}
$i\leq k\leq n: \quad A^i(n) {A^\dg}^k(n) = {A^\dg}^{k-i}(n)$ \, + \, terms with  $A(n)$ to the  right,\\
$i\geq k\leq n: \quad A^i(n) {A^\dg}^k(n) = A^{i-k}(n)$ \, + \, terms with  $A(n)$ to the  right.
\et
\eeq
More explicitly, for $i=1 < k=3$ and $n=3$ the first eq. in (\ref{*3}) reads 
$$ A(3) {A^\dg}^3(3) = {A^\dg}^2(3) -  {A^\dg}^3(3)A(3)[1- {A^\dg}^3(3)A^3(3)] \, + {A^\dg}^3(3)A^2(3){A^\dg}^3(3)A^2(3)[1-{A^\dg}^3(3)A^3(3)],$$
and for $i > k=2$  and any $n$ the second eq. in (\ref{*3}) produces
$$ A^i(n){A^\dg}^2(n) = A^{i-2}(n)[1 - {A^\dg}^n(n) A^n(n)] - A^{i-1}(n){A^\dg}^n(n) A^{n-1}(n)[1-{A^\dg}^n(n) A^n(n)].$$ 
From (\ref{*3}) it follows that 
\begin{equation}\label{A^iA^dg k |0> 2}
A^i(n){A^\dg}^k(n)|0\ra = \left|\begin{tabular}{l}
$0$, \quad if $i>k$,\\
$ |k-i\ra$,\quad if $i \leq k$ .
\end{tabular}\right.
\end{equation} 
Now we readily see that the orthonormality relations (\ref{<i|k>}) follow from (\ref{A^iA^dg k |0> 2}), and
 $A(n)$, $A^\dg(n)$ act on $|k\ra$ as raising and lowering operators with step $1$: 
\begin{equation}\label{A|k>}
A(n)|k\ra =  A(n){A^\dg}^k(n)|0\ra =  |k-1\ra, \quad A^\dg(n) |k \ra = |k+1\ra. 
\end{equation}
Next we have to construct the appropriate particle number operator. 
One can easily check  that $|k\ra$, $k=0,1,\ldots,n$, are eigenstates of the following Hermitian operator  
\begin{equation}\label{N}
N(n) = A^\dg(n)A(n) + {A^\dg}^2(n)A^2(n) + \ldots + {A^\dg}^n(n)A^n(n),  
\end{equation}
the eigenvalues being equal to $k$, 
\begin{equation}\label{N|k>}
N(n)|k\ra = k|k\ra.   
\end{equation}
One easily finds that the squared operator $N^2$  takes the form 
\begin{equation}\label{N^2}
N^2(n) = A^\dg(n)A(n) + 3{A^\dg}^2(n)A^2(n) + 5{A^\dg}^3(n)A^3(n) \ldots + (2n-1){A^\dg}^n(n)A^n(n),  
\end{equation}
wherefrom it follows that $N^2|k\ra = \sum_{i=1}^k(2i-1)|k\ra =  k^2|k\ra$ as required.  
\noi 
Thus $N(n)$ plays the role of the {\it number operator} and states $|k\ra$ can be called {\it number states}. 
The linear span of $|k\ra$  should be denoted as ${\cal H}_{n+1}$, its dimension being $n+1$.   
This is the Fock space for the $n$-fermions. In this space the operator $A^{n+1}(n)$ is to be put to nil, since it annihilates all vectors $|k\ra$:  indeed, from (\ref{A^iA^dg k |0> 2}) we find $A^{n+1}(n)|k\ra = A^{n+1}(n) {A^\dg}^k|0\ra = 0$. It is in this sense that $A(n)$ is nilpotent of order $n+1$, $A^{n+1}(n)=0$.\, 
The set of projectors $|i\ra \la i|$ resolves the identity operator in ${\cal H}_{n+1}$, $\sum_{i}|i\ra\la i| =1$. 
%
In this way the state $|k\ra$, eq. (\ref{|k>}), can be regarded as a normalized state with $k$ number of $n$-fermions, $k=0,1,\ldots,n$. 
There are no states with more than $n$ such particles. 
So the degree of nonlinearity $n$ is the {\it order of statistics}  of our $n$-fermions. 
Let us note that the relation $A(n)A^\dg(n)|0\ra =  |0\ra$ is valid for any $n$ (any order of statistics), while  for the Green parafermions \cite{Green} and Palev $A_n$-type parafermions \cite{Palev} the action of the product of  annihilation and creation operators on $|0\ra$  reads $f\, f^\dg |0\ra = p|0\ra$,   where $p$ is the order of corresponding statistics. 
\medskip

Let us finally find  the commutators between $N(n)$ and $A(n)$ and $A^\dg(n)$. By direct calculations  we get the relations
\begin{equation}\label{[A,N]}
[A(n),N(n)] = A(n), \quad [A^\dg(n),N(n)] = -A^\dg(n),    
\end{equation}
We say that the three operators $A(n),A^\dg(n)$ and $N$, satisfying (\ref{{A,A^dg}}) and (\ref{[A,N]}) form the {\it $n$-fermion algebra}. 
At $n=1$ it coincides with the (standard) fermion algebra. 
 
It is worth noting that the operators $A(n)$, $A^\dg(n)$ can be expressed in terms of the (nonorthogonal) projectors $|i\ra\la k|$ as follows
\begin{equation}\label{|i><k|}
A(n) = \sum_{i=0}^{n-1}|i\ra\la i+1|,\quad A^\dg(n) = \sum_{i=0}^{n-1}|i+1\ra\la i|.  
\end{equation}
The latter formulas show that the operators  $A(n)$, $A^\dg(n)$ can be realized as ladder operators in any finite level quantum system with $n+1$ orthonormalized states. 
This is in complete analogy to the case of ordinary fermion operators and ladder operators in two level systems. 
Matrix elements of $A(n)$ between $|i\ra$ and $|k\ra$ are $\la i|A(n)|k\ra  = \delta_{i,k-1}$.   
 Thus in matrix form we have ($i$ denoting the row, and $k$ - the column)
 
 \begin{equation}\label{A_{ik}}
 A(n)  = \left( \matrix{0&1&0&0&\ldots&0&0\cr 0&0&1&0&\ldots&0&0\cr
                       0&0&0&1&\ldots&0&0\cr
                      .&.&.&.&\ldots&.&.\cr 0&0&0&0&\ldots&0&1\cr
                       0&0&0&0&\ldots&0&0 } \right),    \quad  
  |0\ra = \left( \matrix{1\cr 0\cr  .\cr .\cr .\cr  0 } \right), \quad |n\ra = 
 \left( \matrix{0\cr 0\cr  .\cr .\cr .\cr  1 } \right) .  
 \end{equation} 
\medskip

\section{$n$-Fermion coherent states} 

\subsection{Ladder operator eigenstates}

Our next aim is the construction of coherent states (CS) of $n$-fermion system in the form of annihilation operator eigenstates.
 From $A^{n+1}=0$ it follows that the eigenvalues $\zeta$ of $A$, when exist, should be nilpotent of the same order $n$ as $A$ is, i.e. $\zeta^{n+1}=0$.   
In order to perform the explicit construction of CS we have to specify the (anti-) commutation relations between $\zeta$ and $\zeta^*$, $A$, and $A^\dg$. 
Keeping in mind the known case of  eigenvalues of the ordinary fermion operators we adopt the following relations,
 \begin{equation} \label{zeta A-comm}
\{\zeta,\zeta^*\} = 0,\quad  \{\zeta, A(n)\} = 0 = \{\zeta, A^\dg(n)\} , \quad \zeta |0\ra = |0\ra \zeta,
\end{equation} 
where  $\{a,b\} =ab+ba$. In view of (\ref{zeta A-comm}) and (\ref{|k>}) one has $\zeta |k\ra = (-1)^k|k\ra \zeta$ and $\zeta^*\zeta |k\ra = |k\ra \zeta^*\zeta$.  

The one-mode (complex) Grassmann variables are known to admit matrix representation in terms of $4\times 4$ matrices. It is worth looking therefore for matrix representation of para-Grassmann variables too. In this aim consider the $(n+1)\times(n+1)$ matrices, $n>1$,     
\beq\label{z-algebra}
\uln{\zeta} = \left(\matrix{0 & 0 & 0 & 0& \ldots \cr 1 & 0 & 0 & 0&\ldots\cr 0 & -i & 0 & 0& \ldots \cr 0 & 0 & 1 & 0
&\cdot\cdot\cdot \cr \cdot  & \cdot & \cdot & \cdot& \cdot\cdot\cdot }  \right),\qquad  
\uln{\zeta}^* = \left(\matrix{0 & 0 & 0 & 0& \ldots \cr 1 & 0 & 0 & 0&\ldots\cr 0 & i & 0 & 0& \ldots \cr 0 & 0 & 1 & 0
&\cdot\cdot\cdot \cr \cdot  & \cdot & \cdot & \cdot& \ldots }  \right),
\eeq
where the values $1,\,-i$  are alternating in the first subdiagonal of $\uln{\zeta}$ up to the $n+1$ row, and $\uln{\zeta}^*$ is the congugate of $\uln{\zeta}$.  One can easily verify that the following relations
  are valid:  $\{\uln{\zeta}, \uln{\zeta}^*\} =0$, $\uln{\zeta}^{n+1} = 0 = {\uln{\zeta}^*}^{n+1}$. Therefore these quantities  $\uln{\zeta}$ and $\uln{\zeta}^*$ can be regarded as generators (basis elements) of our para-Grassman algebra, and $z \uln{\zeta}$, $z^*\uln{\zeta}^* $ with $z\in C$ as matrix representation of para-Grassman variables  $\zeta$ and $\zeta^*$.  
\medskip
  
Eigenstates $|\zeta;n\ra$ of $A(n)$ should be constructed as series in terms of the basic number states $|k\ra$. Since $\zeta$ does not commute with  $|k\ra$ it is clear that $\zeta$ could not commute with $|\zeta;n\ra$ too. Therefore there are  two main types of eigenstates of $A(n)$: 'left' and 'right'.  Here we adopt the 'right' eigenstates \footnote{The 'left' eigenstates of $A(n)$ are considered in the last section. At $n=1$ (standard fermion case) the 'left' and 'right' eigenstates coincide.}, 

\begin{equation} \label{A|zeta>_r}
A(n)|\zeta;n\ra_r =  |\zeta;n\ra_r \zeta . 
\end{equation}
Using anticommutation relations (\ref{zeta A-comm}) and the action of $A$ on the number states $|k\ra$ (eq. (\ref{A|k>})) one can easily verify that the superpositions 
\begin{eqnarray}\label{||zeta>_r}
 ||\zeta;n\ra_r &=& |0\ra -\zeta|1\ra + \zeta^2|2\ra - \ldots + (-1)^n\zeta^n|n\ra  \\
  &=& |0\ra +|1\ra \zeta + |2\ra \zeta^2 + \ldots + |n\ra \zeta^n \nonumber
\end{eqnarray}
are (non-normalized) 'right' eigenstates of $A$: \, $A(n)||\zeta;n\ra_r = ||\zeta;n\ra_r\zeta$. The normalization factor ${\cal N}$ turned out to be $\sqrt{1-\zeta^*\zeta}$ so that the normalized states are 
\begin{equation} \label{|zeta>_r}
 |\zeta;n\ra_r = {\cal N}(\zeta^*\zeta) ||\zeta;n\ra_r =  \sqrt{1-\zeta^*\zeta}\sum_{k=0}^n (-\zeta)^k|k\ra.
\end{equation}
At $n=1$ we have $|\zeta;1\ra_r = (1-\zeta^*\zeta/2) \left(|0\ra - \zeta |1\ra\right)$ which recovers the usually used form of standard fermion CS \cite{Cahill}. For any $n$  our states reveal the following useful properties:  
\bea\label{A^k|zeta>_r}
A^k(n) |\zeta;n\ra_r =  |\zeta;n\ra_r \zeta^k,\\
\,_r\la n;\zeta|{A^\dg}^j(n) A^k(n) |\zeta;n\ra_r = {\zeta^*}^j \zeta^k. \label{<A^jA^k|zeta>_r}
\eea
Eigenstates with different eigenvalues $\zeta$ and $\eta$ are not orthogonal. 
Their scalar product takes the form (note that $\zeta\eta=\eta\zeta$, $\zeta\eta^* = -\eta^*\zeta$)
\begin{equation}\label{<zeta||eta>}
\,_r\la n;\zeta|\eta;n\ra_r = \left(1 +\zeta^*\eta + \ldots + {\zeta^*}^n \eta^n\right){\cal N}(\zeta^*\zeta){\cal N}(\eta^*\eta). 
\end{equation}

\subsection{Integration rules and overcompleteness}

Our next aim is to establish the overcompleteness of the set of $A(n)$-eigenstates $ |\zeta;n\ra_r$, eqs. (\ref{||zeta>_r}), (\ref{|zeta>_r}). 
In this purpose we have to find appropreate {\it integration rules for the anticommuting ($n+1$)-order nilpotent  variables}  $\zeta$ and $\zeta^*$, such that the set of projectors $ |\zeta;n\ra_r\,_r\la n;\zeta|$ resolves the unity operator $1$ in the state space ${\cal H}_{n+1}$,
\begin{equation}\label{res 1}
I \equiv \int d\zeta^* d\zeta\,  |\zeta;n\ra_r\,_r\la n;\zeta| = 1.
\end{equation} 
Keeping in mind the  Berezin integration rules for complex Grassmann variables  and the overcompleteness of standard fermion CS \cite{Cahill}  we adopt the following form
 \begin{equation}\label{intrul 1}
\int d\zeta^* d\zeta \, \zeta^i{\zeta^*}^k =  \delta_{ik} g_k(n), 
\end{equation} 
where  $g_k(n)$ are real numbers to be determined.
 We shall require that at $n=1$ the generalized rules (\ref{intrul 1}) recover the Berezin  rules \cite{Cahill}, which read \noi 
\begin{equation}\label{intrul Ber}
g_0^{Ber} =0,\quad g_1^{Ber} = 1. 
\end{equation} 
\noi
In fact the integration $\int d\zeta^*d\zeta$ in (\ref{intrul 1}), and later on, should be regarded {\it as a linear  functional} $I[f]$ that maps functions $f$ of $\zeta$ and $\zeta^*$ into the real line. Due to the nilpotency of $\zeta$ these functions are superpositions of monomials $\zeta^i{\zeta^*}^k$, $i,k=0,1,\ldots, n$.  

\noi Replacing $ |\zeta;n\ra_r$ and $_r\la n;\zeta|$ with the corresponding expressions according to (\ref{|zeta>_r}) we get  
\begin{eqnarray}\label{res 1b}
I = \sum_{i,k=0}^{n}\int d\zeta^*d\zeta\, \sqrt{1-\zeta^*\zeta}\, |i\ra\zeta^{i} {\zeta^*}^{k}\la k| \sqrt{1-\zeta^*\zeta}, 
\end{eqnarray}
wherefrom, taking into account (\ref{intrul 1}),  we have  
\begin{eqnarray}\label{I}
I = \sum_{i,k} \int d\zeta^*d\zeta\, \sqrt{1-\zeta^*\zeta}|i\ra \zeta^{i} {\zeta^*}^{k}\la k| \sqrt{1-\zeta^*\zeta}  = \sum_k I_k,\\
I_k =  \int d\zeta^*d\zeta\, \sqrt{1-\zeta^*\zeta}|k\ra \zeta^{k} {\zeta^*}^{k}\la k| \sqrt{1-\zeta^*\zeta}.  
\end{eqnarray}
 
\noi Noting that $\zeta^{k} {\zeta^*}^{k}$ commute with $|k\ra$ and ${\cal N}$ commutes with $|k\ra\la k|$ and with ${\zeta^*}^k \zeta^k$ we find
\beq\lb{I_k 2}
I_k = \int d\zeta^*d\zeta\, (1-\zeta^*\zeta)\zeta^{k} {\zeta^*}^{k} |k\ra \la k| = (g_k(n) - (-1)^{k+1}g_{k+1})|k\ra\la k|,
\eeq
afterwhat eq. (\ref{I}) is rewritten as
\beq\lb{I 2}
I = \sum_{k=0}^n \left(g_k(n) - (-1)^{k+1}g_{k+1}\right) |k\ra\la k|.
\eeq
In view of the completeness of the set $\{|k\ra\}$ the operator $I$ would be the identity operator in ${\cal H}_{n+1}$ iff 
\beq\lb{recc 1}
 g_k(n) - (-1)^{k+1}g_{k+1} = 1.
\eeq
These are recurrence relations for the quantities $g_k(n)$: one can express all $g_k$ in terms of $g_n$ for example. We put $g_n(n)=1$ and write the solution
\beq\lb{g_k} 
g_k(n) = 1+\sum_{i=1}^{n-k}(-1)^{ki +\frac{i(i+1)}{2}}, \,\, k = n-1,n-2, \ldots,0.
\eeq
For $g_k,\,\, k=n,n-1,n-2,n-3$ we have 
\beq\lb{g_n,g_(n-3)} 
g_n(n) =1,\,\,\, g_{n-1}(n) = 1 + (-1)^n,\,\,\,  g_{n-2}(n) = (-1)^{n-1},\,\,\,  g_{n-3}(n) = 0.
\eeq
Note the {\it different structure} of $g_k(n)$ for odd and even $n$ (i.e. for even and odd  dimension of the space ${\cal H}_{n+1}$).     For odd $n$ ($n=1,3,\ldots$) the structure is very simple:
\beq\lb{odd n}
g_n=1,\,\,\, g_{n-1}=0,\,\,\, g_{n-2} = 1,\,\,\,  \ldots,\,\,\,  g_1=1, \,\,\, g_0=0.
\eeq
Clearly, these are direct and natural generalizations of the Berezin rules (\ref{intrul Ber}) to the case of $n$-order nilpotent anticommuting variables. 

We have proven in the above that the integration rules  (\ref{intrul 1}) with $g_k$ given by (\ref{g_k})  ensure the overcompleteness relation (\ref{res 1}) for $ |\zeta;n\ra_r$. Therefore $ |\zeta;n\ra_r$ can be qualified as CS -- the {\it $n$-fermion ladder operator CS}. At $n=1$ they reproduce the fermionic CS  \cite{Cahill, Abe} (Grassmann CS \cite{Grigorescu}).


\section{General ($n\!+\!1$)-order nilpotent ladder operators and CS}

\subsection{Ladder operators and their 'right' eigenstates}

In the finite dimensional Hilbert space ${\cal H}_{n+1}$  (spanned by a set of orthonormal vectors $|k\ra$) the general form of ($n+1$)-order nilpotent ladder operators that perform transitions between adjacent states are
\beq\lb{A_alf}
A(n,\vec{\alf}) = \sum_{k=0}^{n-1} \alf_k |k\ra\la k+1|,\quad A^\dg(n,\vec{\alf}) = \sum_{k=0}^{n-1} \alf_k^* |k+1\ra\la k|,
\eeq
where $\alf_k$ are complex numbers, in general. The actions on $|k\ra$ are  
\beq\lb{A_alf |k>}
\bt{l}
$\dsty A(n,\vec{\alf})|k\ra = \alf_{k-1}|k-1\ra, \quad A^\dg (n,\vec{\alf})|k\ra = \alf_{k}^* |k+1\ra,$\\[2mm]
$\dsty A^\dg (n,\vec{\alf})A(n,\vec{\alf})|k\ra = |\alf_{k-1}|^2|k\ra.$
\et
\eeq
We suppose that $A(n,\vec{\alf})|k\ra|0\ra = 0$, so that $\alf_{-1}=0$. States $|k\ra$ may be regarded as eigenstates of a Hamiltonian $H$ of some finite level ($n+1$ levels) quantum system, not necessarily  with equidistant spectrum.       If $\veps_0=0,\, \veps_1,\ldots,\veps_n$ are (nondegenerate) eigenvalues of $H$ then we have $H=A^\dg (n,\vec{\alf})A(n,\vec{\alf})$ with $|\alf_{k-1}|^2 = \veps_k$. Since $|k\ra$ are eigenstates also of  $n$-fermion number operator $N$,  eq. (\ref{N}), the Hamiltonian commutes with $N$. The proportionality relation $H=\hbar\omega N$ ($\omega$ being some frequency parameter) holds for $H$ with equidistant specrum only. Thus  $\veps_k$ may be viewed as a sum of energy of $k$ noninteracting $n$-fermions with equal portions $\veps_k/k$, $k=1,2,\ldots,n$.   

In previous sections we studied the simplest case of $\alf_k=1$, i.e. we have $A(n,1) \equiv A(n)$.  This means that  nonlinear fermion operators $A(n)$ can be formally realized for any $n+1$ quantum system, in complete analogy to the case of standard fermion operators and $2$-level system. At $\alf_k = \sqrt{(k+1)(2j-k)}$ the spin $j$ lowering operators $J_-$ are recovered (redenoting standard spin states $|j,m\ra$ with $|k\ra$, $k = m+j$: \,\, $|j.-j\ra = |0\ra$, $|j,-j+1\ra = |1\ra, \,\, \ldots, \,\, |j,j\ra = |n\ra$, $n=2j$; \,\, $\alf_{-1}=0,\, \alf_0 = \sqrt{n}, \, \alf_{1}= \sqrt{2(n-1)}\ldots, \alf_{n-1} =\sqrt{n}$ ) .  

As in the known case of standard fermions we can consider the $n$-fermion operators $A(n)$, $A^\dg(n)$ as a complete set in the sense that any other operator in ${\cal H}_{n+1}$ is a polynomial in terms of $A(n)$, $A^\dg(n)$.  It is not difficult to verify that the general ladder operators $A(n,\vec{\alf})$ take the following polynomial form:  
\beq\lb{A_alf- A_1}
A(n,\vec{\alf}) = \alf_0 A(n) + (\alf_1-\alf_0)A^\dg(n)A^2(n) +\ldots (\alf_{n-1}-\alf_{n-2}){A^\dg}^{n-1}(n)A^n(n).
\eeq 
 In the particular case of spin ladder operators $J_\pm$ for spin $j=n/2=1/2,1,3/2$ such formula is provided in \cite{Demichev}. Clearly, in general the ladder operators (\ref{A_alf}) do not obey the nonlinear anticommutation relation (\ref{{A,A^dg}}), though the operator 
$A(n,\vec{\alf})A^\dg(n,\vec{\alf}) + {A^\dg}^n(n,\vec{\alf})A^n(n,\vec{\alf})$ is  diagonal: 
\beq\lb{diag}
A(n,\vec{\alf})A^\dg(n,\vec{\alf}) + {A^\dg}^n(n,\vec{\alf})A^n(n,\vec{\alf}) = {\rm diag}\{|\alf_0|^2, |\alf_1|^2, \ldots, |\alf_{n-1}|^2, \,  |\alf_{n-1}|^2!\},
\eeq
where $|\alf_{n-1}|^2! = |\alf_0|^2 |\alf_1|^2 \ldots |\alf_{n-1}|^2$.

In the aim to construct CS as eigenstates of $A(n,\vec{\alf})$ we first note that the eigenvalues $\zeta$ of  $n$-fermion annihilation operators $A(n)$ anticommute with $A(n,\vec{\alf})$, as it is seen from eqs. (\ref{A_alf- A_1}), (\ref{zeta A-comm}): 
\beq\lb{{A_alf,z}}
\{\zeta, A(n,\vec{\alf})\} = 0 = \{\zeta, A^\dg(n,\vec{\alf})\}.
\eeq
Then, using (\ref{A_alf |k>}) and (\ref{{A_alf,z}}),  we easily derive the form of eigenstates of $A(n,\vec{\alf})$ with 'right' eigenvalue (shortly 'right' eigenstates),
\beq\lb{|zeta;alf>}
A(n,\vec{\alf}) |\zeta;n,\vec{\alf}\ra_r = |\zeta;n,\vec{\alf}\ra_r \zeta, \quad 
|\zeta\ra_r = {\cal N}(\zeta^*\zeta,\vec{\alf}) ||\zeta;n,\vec{\alf}\ra_r,
\eeq
\begin{eqnarray}\label{||zeta;alf>}
 ||\zeta;n,\vec{\alf}\ra_r &=& |0\ra -\frac{\zeta}{\alf_0!}|1\ra + \frac{\zeta^2}{\alf_1!}|2\ra - \frac{\zeta^3}{\alf_2!}|3\ra + 
\ldots + (-1)^n\frac{\zeta^n}{\alf_{n-1}!}|n\ra    \nonumber  \\
  &=& \sum_{k=0}^n  (-1)^k  \frac{ \zeta^{k}}{\alf_{k-1}!}|k\ra, 
\end{eqnarray}
where $\alf_{-1} \equiv  0$, $0!=1$, $\alf_{k-1}! =  \alf_0\alf_1\ldots\alf_{k-1}$. \,\, ${\cal N}(\zeta^*\zeta,\vec{\alf})$ is the normalization constant, 
\beq\lb{N_alf}
{\cal N}^2(\zeta^*\zeta,\vec{\alf}) = \sum_{j=0}^n a'_j(\vec{\alf})(\zeta^*\zeta)^j 
=  \sum_{j=1}^n a_j(\vec{\alf}){\zeta^*}^j\zeta^j,
\eeq
the coefficients $a_j$ being determined via the recurrence relations
\beq\lb{a_k}
a_j = -\frac{a_0}{|\alf_j|^2!} - \frac{a_1}{|\alf_{j-1}|^2!} - \ldots - \frac{a_{j-1}}{|\alf_1|^2!}.
\eeq   
The first four $a_j$ are $a_0$, $a_1 = -a_0/|\alf_1|^2!$,  $a_2 = -a_0/|\alf_2|^2! - a_1/|\alf_{1}|^2!$, $a_3 =  -a_0/|\alf_3|^2! - a_1/|\alf_{2}|^2! - a_2/|\alf_{1}|^2!$, which in terms of $\alf_k$ read ($a_0$ is put to $1$)
\beq
\bt{l}\lb{a0-a3}
$\dsty a_0=1,\quad a_1 = -\frac{1}{|\alf_1|^2!}$,\\[3mm]
$\dsty a_2 = -\frac{1}{|\alf_2|^2!} + \frac{1}{|\alf_1|^4!}$,\\[3mm]
$\dsty a_3 = -\frac{1}{|\alf_3|^2!} + \frac{2}{|\alf_1|^2!|\alf_2|^2!} - \frac{1}{|\alf_1|^6!}$.
\et
\eeq

\medskip

\subsection{Overcompleteness of the 'right' eigenstates of $A(n,\vec{\alf})$}

Our next task is the overcompleteness of the set $\{|\zeta;n,\vec{\alf}\ra_r\}$.  It turns out that the para-Grassmann variables $\zeta,\,\zeta^*$ with the same {\it basic anticommutation and integration rules} (eqs. (\ref{zeta A-comm}), (\ref{intrul 1}), (\ref{{A_alf,z}}) and (\ref{g_k})) can be used to establish the overcompleteness of this  set.  This time however  we have to introduce appropreate weight function $W(\zeta^*\zeta;n,\vec{\alf})$ in the resolution unity relation,
\begin{equation}  \label{res 1, alf}
\int d\zeta^*d\zeta\,W(\zeta^*\zeta;n,\vec{\alf}) |\zeta;n,\vec{\alf}\ra_r \, _r\la \vec{\alf},n;\zeta|  = 1. 
\end{equation}  
Taking into account that $W$ and ${\cal N}$ depend on $\zeta$ and $\zeta^*$ through $\zeta^*\zeta$ and using eqs.  (\ref{||zeta;alf>}),  (\ref{intrul 1}),  and the completeness of the set of $|k\ra$,  we obtain the $n+1$ integral relations for the unknown $W$, 
\begin{equation}  \label{ints W k}
\int d\zeta^* d\zeta\,W(\zeta^*\zeta;n,\vec{\alf}) {\cal N}^2(\zeta^*\zeta;n,\vec{\alf})\,  
\frac{\zeta^k  {\zeta^*}^k}{|\alf_{k-1}|^2!} = 1, 
\end{equation} 
where $k=0,1,\ldots,n,$, $|\alf_{k-1}|^2! = |\alf_0|^2|\alf_1|^2 \ldots |\alf_{k-1}|^2$ and $|\alf_{-1}|^2! = 0!=1$. \, In fact here we have to solve the inverse momentum problem.  We expand $W$ in a series in terms of $\zeta^*\zeta$,
\begin{equation} \label{W sum}
W(\zeta\zeta^*;n,\vec{\alf}) = \sum_{i=0}^n w'_i(n,\vec{\alf})\,(\zeta \zeta^*)^i 
= \sum_{i=0}^n w_i(n,\vec{\alf})\,{\zeta^*}^i \zeta^i,   
\end{equation} 
take into account the expansion of ${\cal N}^2$, eq. (\ref{N_alf}),  and get 
\begin{equation}  \label{ints W k 2}
\frac{1}{|\alf_{k-1}|^2!} 
\sum_{i=0,j=0}^n a_jw_i \int d\zeta^*d\zeta\,{\zeta^*}^i\zeta^i\,{\zeta^*}^j \zeta^j\, \zeta^k  {\zeta^*}^k = 1. 
\end{equation} 
In view of (\ref{intrul 1}) these are  recurrence relations  for $w_i$ in terms of the known $\alf_{k-1}$, $a_k$  and $g_k$. Indeed, for $k=n$ in (\ref{ints W k 2}) we have $a_0w_0 g_n = |\alf_{n-1}|^2!$, wherefrom $w_0$ is determined. Similarly putting $k=n-1, n-2, \ldots $ we can express $w_1,\, w_2,\, \ldots $ in terms of $w_0,\,w_1,\, \ldots$,
\beq\lb{w0-w3}
\bt{ll}
$\dsty w_0 = $&$\dsty \frac{|\alf_{n-1}|^2!}{a_0g_n},$ \\
$\dsty w_1 = $&$\dsty (-1)^{n}\frac{1}{a_0g_n}\left( |\alf_{n-2}|^2!  - a_0w_0 g_{n-1}  - (-1)^{n}a_1w_0 g_n\right),$\\
$\dsty w_2 = $&$\dsty \frac{1}{a_0g_n}\left(|\alf_{n-3}|^2! - a_0w_0g_{n-2} - (-1)^{n-1}(a_0w_1+a_1w_0)g_{n-1}\right)$\\[2mm]
         $\dsty        \,\,     $&$\dsty  \quad + \, \frac{1}{a_0}\left(a_1w_1-a_2w_0\right),$\\
$\dsty w_3  = $&$\dsty (-1)^{3n}\frac{1}{a_0g_n}\left(|\alf_{n-4}|^2! - a_0w_0g_{n-3}
 -  (-1)^{n-2}(a_0w_1+a_1w_0)g_{n-2}\right)$\\[2mm]
 $\dsty             $&$\dsty  -  \frac{1}{a_0}\left( (-1)^{3n} (a_0w_2-a_2w_0-a_1w_1)g_{n-1})  -(a_3w_0-w_1a_2-w_2a_1)\right).$
\et
\eeq
At $\alf_k=1$ the above formulas (\ref{w0-w3}) do reproduce $w_0=1$, $w_1=0=w_2=w_3$ as expected.  
We write down the explicit form of all $W$-coefficients $w_i$ in terms of $\alf_k$ for $n=3$ ($4$-level system):
\beq\lb{w_i n3}
\bt{ll}
$ w_0 =$& $|\alf_{2}|^2!,\qquad w_1 = - |\alf_{1}|^2!   - a_1|\alf_2|^2!,$ \lb{w1 n3}\\[2mm]
$ w_2 =$&$  |\alf_0|^2!   - |\alf_2|^2! - a_1\left(|\alf_1|^2! +a_1|\alf_2|^2!\right) - a_2|\alf_2|^2!,$\\[2mm]
$w_3 =$&$  -1 + |\alf_1|^2!   + a_3 |\alf_2|^2!  + a_2\left(|\alf_1|^2! +a_1|\alf_2|^2!\right)$ \\[1mm]
$        \,\,$&$  - a_1\left(|\alf_0|^2! - |\alf_2|^2! - a_1|\alf_1|^2! - a_1^2|\alf_2|^2! - a_2|\alf_2|^2!\right)$,  
\et\eeq
where $a_j$ are given in terms of $\alf_k$ in (\ref{a0-a3}), $a_0=1$.

It worths emphasizing that the resolution of unity operator in terms of the non-normalized eigenstates projectors $||\zeta;\vec{\alf},n\ra_r \, _r\la n,\vec{\alf};\zeta||$, eq. (\ref{||zeta>_r}), is also possible,
\begin{equation}  \label{res 2, alf }
\int d\zeta^*d\zeta\,\widetilde{W}(\zeta^*\zeta;n,\vec{\alf}) ||\zeta;n,\vec{\alf}\ra_r \, _r\la \vec{\alf},n;\zeta||  = 1, 
\end{equation}  
with the new weight function  $\widetilde{W}(\zeta^*\zeta;n,\vec{\alf})$,
\begin{equation} \label{tld W sum}
\widetilde{W}(\zeta^*\zeta;n,\vec{\alf}) = \sum_{i=0}^n \tilde{w}_i(n,\vec{\alf})\,{\zeta^*}^i \zeta^i,   
\end{equation} 
the coefficients $\tilde{w}_i$ being determined via the recurrence relations (\ref{ints W k 2}), this time with $a_0=1$ and $a_1=a_2 = \ldots = a_n =0$. For $n=3$ ($4$-level system) the four $\tilde{w}_i$  read
\beq\lb{tld w_i n3}
\bt{ll}
$ \tilde{w}_0 =$& $|\alf_{2}|^2!,\qquad \tilde{w}_1 = - |\alf_{1}|^2!$ \\[2mm]
$ \tilde{w}_2 =$&$  |\alf_0|^2!   - |\alf_2|^2! ,\quad  \tilde{w}_3 =  -1 + |\alf_1|^2!$.  
\et\eeq  
At $\alf_k=1$ the relations (\ref{tld w_i n3}) produce  $\widetilde{W} = 1-\zeta^*\zeta$ as expected.
\medskip 

In this way we have proved that the ladder operator 'right' eigenstates $|\zeta;n,\vec{\alf}\ra_r$ for every set of nonvanishig complex parameters $\alf_k$ do form an overcomplete family with respect to anticommutation relations (\ref{zeta A-comm}) and  the fixed integration rules (\ref{intrul 1}), (\ref{g_k}). Therefore these states can be qualified as CS ({\it  ladder operator 'right' CS} ) for the ($n+1$)-level quantum systems.        At $\alf_k = \sqrt{(k+1)(2j-k)}$ the operator $A(n,\vec{\alf})$ coincides with spin lowering  operator $J_-$, therefore the states $|\zeta;n,\vec{\alf}\ra_r$ at  $\alf_k = \sqrt{(k+1)(2j-k)}$ are spin ladder operator 'right' CS.  At $\alpha_k = 1$ for all $k$ the states $|\zeta;n,\vec{\alf}\ra_r$  reproduce the $n$-fermion 'right' CS $|\zeta;n\ra_r$. \,  
In CS $|\zeta;n,\vec{\alf}\ra_r$ the eigenvalues of $A^k(n,\vec{\alf})$ and the mean values of ${A^\dg}^i(n,\vec{\alf}) A^k(n,\vec{\alf})$, for any $n$ and $\vec{\alf}$ , are given by the same desired expressions as in (\ref{A^k|zeta>_r}), (\ref{<A^jA^k|zeta>_r}).   


\section{'Left' ladder operator CS    and displacement-operator-like CS  } 

\subsection{ 'Left' ladder operator CS}

Our aim here is the construction of ladder operator 'left' CS of {\it general} ($n+1$)-level system with orthonormalized states $|k\ra$. As in the previous sections we suppose that  eigenvalues $\zeta$ and the ladder operators $A(n,\vec{\alf})$, eq. (\ref{A_alf}), obey  the same (anti-)commutation relations (\ref{zeta A-comm}). Then we easily find that the states 
\begin{eqnarray}\label{||zeta>_l}
 ||\zeta;n,\vec{\alf})\ra_l &=& \sum_{k=0}^n (-1)^{[\frac{k+1}{2}]} \frac{\zeta^k}{\alf_{k-1}!}|k\ra \\
  &=& |0\ra - \frac{\zeta}{\alf_0!}|1\ra - \frac{\zeta^2}{\alf_1!}|2\ra + \frac{\zeta^3}{\alf_2!}|3\ra + 
\ldots + (-1)^{[\frac{n+1}{2}]} \frac{\zeta^n}{\alf_{n-1}!}|n\ra \nonumber
\end{eqnarray}
are (non-normalized)  eigenstates of $A(n,\vec{\alf})$ with 'left' eigenvalue (shortly 'left' eigenstates), 
\begin{equation} \label{A||zeta>_l}
A(n,\vec{\alf}))||\zeta;n,\vec{\alf})\ra_l =  \zeta ||\zeta;n,\vec{\alf})\ra_l. 
\end{equation}
 At $n=1$ and $\alf_k=1$  'left' and 'right' eigenstates coincide:  we have $||\zeta;1,1\ra_l = |0\ra - \zeta |1\ra$ which recovers the usually used form of standard fermion CS \cite{Cahill}. 

An objectionable property of these states is that the eigenvalue of even powers of $A(n,\vec{\alf})$ is not equal to the same power of $\zeta$,  
\beq\label{A^k|zeta>_l}
A^k(n,\vec{\alf})||\zeta;n,\vec{\alf}\ra_l = (-1)^{k-1}\zeta^k||\zeta;n,\vec{\alf}\ra_l.
\eeq
The scalar product \,$_l\la\vec{\alf},n;\zeta||\zeta;n, \vec{\alf}\ra_l$ takes the Hermitian form 
\beq\lb{<z||z>}
\, _l\la \vec{\alf},n;\zeta||\zeta;n,\vec{\alf}\ra_l = 1 +\frac{\zeta^*\zeta}{|\alf_0|^2} + \frac{{\zeta^*}^2\zeta^2}{|\alf_1|^2!} + \ldots + \frac{{\zeta^*}^n\zeta^n}{|\alf_{n-1}|^2!}.
\eeq 
The normalization constant  can be seen to be the same as that in eq. (\ref{N_alf}): ${\cal N}_l= {\cal N}$. 
However in view of the noncommutation between $\zeta$ and ${\cal N}$ we encounter another objectionable property:   the normalized 'left' states $|\zeta;n,\vec{\alf}\ra_l := {\cal N}||\zeta;n,\vec{\alf}\ra_l$ are not eigenstates of $A(n,\vec{\alf})$ for $n>1$,
\beq\lb{A|z>_l}
\bt{ll}
$\dsty A(n,\vec{\alf}){\cal N} ||\zeta;n,\vec{\alf}\ra_l $&$ =  \,\, {\cal N} A(n,\vec{\alf})||\zeta;n,\vec{\alf}\ra_l = 
{\cal N} \zeta ||\zeta;n,\vec{\alf}\ra_l $ \\
                           \,\,                                                      & $ \neq  \,\, \zeta {\cal N} ||\zeta;n,\vec{\alf}\ra_l$.
\et
\eeq
As a result  further  objectionable properties occur:   at $n>1$ the meanvalues  in $|\zeta;n,\vec{\alf}\ra_l$ of powers $A^k$ and ${A^\dg}^j A^k$ are not equal to $\zeta^k$ and ${\zeta^*}^j\zeta^k$ respectively. 

It is interesting however that the 'left' eigenstates $|\zeta;n,\vec{\alf}\ra_l $ do form an overcomplete set with respect to our integration rules (\ref{intrul 1}), (\ref{g_k}), the weight function $W(\zeta^*\zeta;n,\vec{\alf})$ being  the same as that for the 'right' CS (see eqs. (\ref{W sum})-(\ref{w_i n3})). This can be verified by direct calculations. So, mathematically, we have two overcomplete families of normalized states related to every ladder operator $A(n,\vec{\alf})$. They coincide at $n=1$ only. 

It is worth noting that    (nonnormalized) eigenstates of operators of the form (\ref{A_alf})  with 'left' eigenvalue are constructed in the recent papers \cite{Cabra'06, Fasihi'10, N-bashi'11, Maleki'11}, where different  anticommutation relations and integration rules for generalized Grassmann variables were adopted. 
These 'left' eigenstates, called in    \cite{Cabra'06, Fasihi'10, N-bashi'11, Maleki'11}   CS, reveal similar objectionable properties      as those for $|\zeta;n,\vec{\alf}\ra_l $, listed above, which are due to the noncommutation of the corresponding paragrassmannian eigenvalues with the related normalization constants and ladder operators (The way out of such inconsistencies is provided by the 'right' CS, constructed in the previous subsections). At $\alpha_k = 1$ for all $k$ the states $|\zeta;n,\vec{\alf}\ra_l$  are 'left' eigenstates of the $n$-fermion annihilation operator $A(n)$, and at  $\alf_k = \sqrt{(k+1)(2j-k)}$ they are 'left' eigenstates of spin lowering operator $J_-$.   

\subsection{ Displacement-operator-like CS for $n$-fermions}

The canonical fermion CS $ |\zeta\ra$ \cite{Cahill}, reproduced by our  CS $|\zeta;n\ra$ at $n=1$,  are equally well constructed as displacement-operator CS  $|\zeta\ra= D(\zeta)|0\ra$ with the unitary desplacement operator $D(\zeta) = \exp(a^\dg \zeta - \zeta^* a)$, $a\equiv A(1)$. These $D(\zeta)$ obey the relation 
\beq\label{1*}
D^\dg(\zeta) \,a\, D(\zeta) =a+\zeta, 
\eeq 
wherefrom it follows, due to the  commutations $[\zeta,\, a^\dg \zeta - \zeta^* a]=0$,  $[a,\,a^\dg \zeta - \zeta^* a)]=\zeta$, that $D(\zeta)|0\ra$ is (left and right) eigenstate of $a$ with eigenvalue $\zeta$. 

One might be tempted to construct displacement-operator CS  for $n$-fermions by means of similarly defined unitary displacement operator 
\beq\label{2*}
D(\zeta,n) =  \exp(A^\dg(n) \zeta - \zeta^* A(n)),
\eeq 
which at $n=1$ recover the fermion displacement operator \cite{Cahill}. However at  $n>1$  the variable $\zeta$ does not commute with $A^\dg(n) \zeta - \zeta^* A(n)$ and  $[A(n),\,A^\dg(n) \zeta - \zeta^* A(n)] \neq \zeta$, which means that  
\beq\label{3*}
D^\dg(\zeta,n) \,A(n)\, D(\zeta,n) \neq  A(n) + \zeta.
\eeq  
Therefore  the 'displaced' states $|\zeta;n\ra_D$, 
\beq\label{|4*}
|\zeta;n\ra_D := D(\zeta,n)|0\ra,
\eeq
{\it are not eigenstates}\, of $A(n)$. They however do form an overcomplete system with respect to our integration rules (\ref{intrul 1}), (\ref{g_k}) and with appropreately determined weight function $W_D(\zeta^*\zeta;n)$. On can easily verify the validity of these two statements on the example of $n=2$, $n =3$ etc.  For $n=2$  we find 
\bea\label{5*} 
|\zeta;2\ra_D &=& \left(1-\frac{1}{2}\zeta^*\zeta\right)|0\ra - 
\zeta|1\ra - \frac{1}{2}\zeta^2|2\ra,\\ 
W_D(\zeta^*\zeta;2) &=& 4 - 7\zeta^*\zeta - 9{\zeta^*}^2\zeta^2, \label{6*} 
\eea
and for $n=3$: 
\bea\label{7*} 
 |\zeta;3\ra_D  &=&  \left(1 - \frac{1}{2}\zeta^*\zeta - \frac{1}{6!} {\zeta^*}^3\zeta^3\right)|0\ra  -  \left(\zeta - \frac{1}{5!}{\zeta^*}^2\zeta^3\right)|1\ra \nonumber\\
 &\quad+& \left(-\frac{1}{2}\zeta^2 + \frac{1}{4!} {\zeta^*}\zeta^3\right)|2\ra 
+  \frac{1}{3!} \zeta^3 |3\ra, \\
W_D(\zeta^*\zeta;3) &=&  36 + 2\zeta^*\zeta   + \frac{178}{5}{\zeta^*}^2\zeta^2 +
\frac{1382}{20} {\zeta^*}^3\zeta^3. \label{8*}   
\eea
Using the nilpotency of $\zeta$ and the (anti)commutation relations (\ref{zeta A-comm}) one can directly check that the states $|\zeta;2\ra_D,\,  |\zeta;3\ra_D$, given by the series in (\ref{5*}), (\ref{7*}), are normalized. In view of the fact that  the normalized 'displaced' states $|\zeta;n\ra_D$ are not eigenstates of $A(n)$ we shall call them  {\it displacement-operator-like} CS ($D$- CS) for $n$-fermions. 
 
It is worth noting that for $n$-fermions there exist  'higher order'  unitary 'displacement-like' operators due to the fact that exponentials of polynomials in terms of $A^\dg(n)$, $A(n)$ are finite series. Applying such operators to the ground state $|0\ra$ one can expect to obtain other overcomplete families of states (CS). For example the states   
\beq\label{9*}
|\zeta;n\ra_{D^\prime} := D^\prime(\zeta,n)\,|0\ra,
\eeq
where
\beq\label{10*}
D^\prime(\zeta,n) = \exp\left( \sum_{k=1}^n {A^{\dg}}^k(n) \zeta^k - {\zeta^*}^k A^k(n)\right),
\eeq 
are normalized and form an overcomplete set with respect to rules (\ref{intrul 1}), (\ref{g_k}) and with appropreately determined weight function $W_{D^\prime}(\zeta^*\zeta;n)$. These $|\zeta;n\ra_{D^\prime}$, like $|\zeta;n\ra_D$, are not eigenstates of $A(n)$. For illustration let us provide the results for $n=2$,
\bea\label{11*} 
|\zeta;2\ra_{D^\prime} &=& \left(1-\frac{1}{2}\zeta^*\zeta - \frac{1}{2}{\zeta^*}^2\zeta^2\right)|0\ra - 
\left(\zeta + \frac{1}{2}\zeta^*\zeta^2\right)|1\ra + \frac{1}{2}\zeta^2|2\ra,\\ 
W_{D^\prime}(\zeta^*\zeta;2) &=& 4 - 11\zeta^*\zeta - 9{\zeta^*}^2\zeta^2, \label{12*} 
\eea
 and for $n=3$, 
\bea\label{13*} 
 |\zeta;3\ra_{D^\prime}  & =&  \left(1 - \frac{1}{2}\zeta^*\zeta - \frac{1}{2}\zeta^*\zeta^2 - \frac{1}{4!}  {\zeta^*}^2\zeta^3 - \frac{21}{5!} {\zeta^*}^3\zeta^3\right)|0\ra \nonumber\\
 &\quad+& \left(-\zeta - \frac{1}{2}\zeta^*\zeta^2 + \frac{1}{3!}\zeta^*\zeta^3 +  \frac{59}{5!} {\zeta^*}^2\zeta^3 + \frac{1}{4!} {\zeta^*}^3\zeta^3\right)|1\ra \nonumber\\
 &\quad+& \left(\frac{1}{2}\zeta^2 - \frac{7}{4!} \zeta^*\zeta ^3+ \frac{1}{3!} {\zeta^*}^2\zeta^3\right)|2\ra 
 + \left( \frac{2}{3}\zeta^3 +  \frac{1}{4!} {\zeta^*}^3\zeta^3 \right)|3\ra, \\
W_{D^\prime}(\zeta^*\zeta;3) &=&  \frac{9}{4} - \frac{11}{8}\zeta^*\zeta   - \frac{9}{10}{\zeta^*}^2\zeta^2 -
\frac{143}{32} {\zeta^*}^3\zeta^3. \label{14*}   
\eea

Thus there are three {\it distinct} types of overcomplete families of $n$-fermion  CS  ('right' and 'left' ladder operator CS), and displacement-operator-like CS ($D$- and $D^\prime$- CS), which at $n=1$ all recover the canonical fermion CS \cite{Cahill}. It is worth recalling at this point the similar situation with the well known Barut-Girardello CS $|z;k\ra$ and Perelomov CS $|\xi,k\ra$ for $SU(1,1)$ (see refs. e.g. in \cite{Klauder}):  the former are eigenstates of the  ladder operator $K_-$ with eigenvalue $z$, and the latter are obtained applying to the ground state the $SU(1,1)$ unitary operator $D(\xi) = \exp(\xi K_+ - \xi^*K_-)$. However $|\xi,k\ra$ are not eigenstates of $K_-$, and $|z;k\ra$ are not orbit of $D(\xi)$.

\section{Concluding remarks}

We have introduced new kind of parafermions,  called $n$-linear fermions (shortly $n$-fermions), based on the simple non-linear anticommutation relations (\ref{{A,A^dg}}). Using these relations and supposing the existence of unique vacuum we {\it derived}      (not simply assumed)    the ($n+1$)-order nilpotency of the operators $A(n),\, A^\dg(n)$ and constructed the system of Fock states, on which $A(n)$ and $A^\dg(n)$ act as particle creation and annihilation operators. The $(n+1)\times(n+1)$ matrix realization of these operators is provided. A remarkable feature of $n$-fermions is that the order of their statistics equals the degree of nonlinearity $n$ of the anticommutation relations (\ref{{A,A^dg}}). 

'Right' and 'left' eigenstates of $n$-fermion annihilation operator $A(n)$  are presented, introducing generalized   (($n+1$)-order nilpotent)   Grassmann variables $\zeta$ and $\zeta^*$ with the same simple anticommutation relations as for standard Grassmann variables.     These para-Grassmann variables are shown to be realized by simple $(n+1)\times(n+1)$ matrices, eq. (\ref{z-algebra})  . Then we derived integration rules for  these variables that ensure the overcompleteness of eigenstates of $A(n)$, establishing in this way $n$-fermion CS. Our integration rules are a simple generalization of Berezin rules used for canonical fermion CS \cite{Cahill}, but differ from rules used in other papers on parafermion CS  \cite{Daoud'02}-\cite{Baz'10}, \cite{Fasihi'10}-\cite{Maleki'11}.  They reveal some differences for odd and even $n$, the rules for  odd $n$ (even dimension of the Fock space) being most simple. This is an interesting correlation between (parity of) paraparticle order of statistics and related para-Grassmann integration rules.  

The $n$-fermion para-Grassmann variables and integration rules appeared in certain sense as basic ones. Using these same variables and rules we succeeded to construct 'right' and 'left' overcomplete ladder operator CS for the general ladder operators $A(n,\vec{\alf})$, $A^\dg(n,\vec{\alf})$ in the 
($n+1$)-dimensional Hilbert space ${\cal H}_{n+1}$,    and normalized displacement-operator-like CS for $n$-fermions  .   At $n=1$ the 'right', 'left'    and the displacement-operator-like CS all   reproduce the canonical fermion CS, but at $n>1$  these CS reveal different properties, more useful in case of 'right' CS and some objectionable ones in case of 'left' CS. The operators $A(n,\vec{\alf})$, $A^\dg(n,\vec{\alf})$ were  expressed as polynomials in terms of $A(n),\, A^\dg(n)$.  It is feasible that any operator in ${\cal H}_{n+1}$ could be expressed in terms of $n$-fermion operators, as it is the case with ordinary fermions at $n=1$.     The noted commutation of the Hamiltonian of a finite level system with the $n$-fermion number operator reveals the possibility to consider the energy level $\veps_k$ as a sum of the energies of $k$ number of $n$-fermions. The concept of $n$-fermions, the related new para-Grassmann algebra and CS could be useful in further description of physical properties of finite level quantum systems.       
\vspace{5mm}

{\large\bf Acknowlegement.} The author thanks the referees for valuable remarks which helped to improve the presentation.

\end{document}